\documentclass[twocolumn,aps,showpacs]{revtex4}
\usepackage{epsfig}
\usepackage{times}

\newcommand{\C}{\emph{C}}
\newcommand{\D}{\emph{D}}
\newcommand{\Lo}{\emph{L}}
\newcommand{\PD}{PD}
\newcommand{\PGG}{PGG}
\newcommand{\MC}{MC}
\newcommand{\DP}{DP}
\newcommand{\x}{$\bf x$}
\newcommand{\y}{$\bf y$}
\newcommand{\sx}{$s(\bf x)$}
\newcommand{\Px}{$P(\bf x)$}
\newcommand{\lfrac}[2]{\frac{\displaystyle #1}{\displaystyle #2}}

\begin{document}

\title{Phase transitions and volunteering in spatial public goods games}
\author{Gy\"orgy Szab\'o}
\affiliation
{Research Institute for Technical Physics and Materials Science
P.O.Box 49, H-1525 Budapest, Hungary}
\author{Christoph Hauert}
\affiliation{Institute for Mathematics, University of Vienna,
Strudlhofgasse 4, A-1090 Vienna, Austria}
\altaffiliation[Present address:]{Department of Zoology, University of British
Columbia, 6270 University Boulevard, Vancouver, B.C., Canada V6T 1Z4}

\date{\today}

\begin{abstract}
Cooperative behavior among unrelated individuals in human and animal societies
represents a most intriguing puzzle to scientists in various disciplines. Here we 
present a simple yet effective mechanism promoting cooperation under full 
anonymity by allowing for voluntary participation in public goods games. This natural
extension leads to rock--scissors--paper type cyclic dominance of the three strategies
cooperate, defect and loner i.e. those unwilling to participate in the public enterprise.
In spatial settings with players arranged on a regular lattice this results in interesting
dynamical properties and intriguing spatio-temporal patterns. In particular, variations
of the value of the public good leads to transitions between one-, two- and 
three-strategy states which are either in the class of directed percolation or show 
interesting analogies to Ising-type models. Although volunteering is incapable 
of stabilizing cooperation, it efficiently prevents successful spreading of selfish 
behavior and enables cooperators to persist at substantial levels.
\end{abstract}

\maketitle

In behavioral sciences and more recently in economics the evolution of cooperation
among unrelated individuals represents one of the most stunning phenomena
\cite{neumann:44,maynard:95}. The prisoner's dilemma (\PD) has long established
as a paradigm to explain cooperative behavior through pairwise interactions
\cite{axelrod:science81}. While the \PD\ attracted attention from biologists and
social scientists, most studies in experimental economics focused on the closely
related but more general public goods game (\PGG) for group interactions
\cite{kagel:95}. In typical \PGG\ experiments, an experimenter endows e.g. 
four players with \$ 10 each. The players then have the opportunity to invest
part or all of their money into a common pool. They know that the total amount
in the pool is doubled and equally divided among all participants irrespective
of their contributions. If everybody cooperates and contributes their money,
each player ends up with \$ 20. However, every player faces the temptation to
defect and to free-ride on the other player's contributions by withholding the
money since every invested dollar returns only 50 cents to the investor.
Obviously, defection represents the dominating strategy leading to the 'rational'
equilibrium where no one increases its initial capital. Such strategical behavior
prescribed to \emph{homo oeconomicus} is frequently at odds with experimental
findings \cite{fehr:nature02} and lead to the decline of this rationality concept.

Note that for pairwise encounters with a fixed investment amount, the \PGG\
reduces to the \PD. \PGG\ interactions are abundant in animal and human societies
\cite{dugatkin:97,colman:95,binmore:94}. Consider for example predator inspection
behavior, alarm calls and group defense as well as health insurance, public
transportation or environmental issues, to name only a few.

Recently it was demonstrated that voluntary participation in such public enterprises
may provide an escape hatch out of economic stalemate and results in a substantial
and persistent willingness to cooperate even in sizable groups, in absence of
repeated interactions, under full anonymity and without secondary mechanisms such
as punishment or reward \cite{hauert:science02}.

The voluntary participation in the \PGG\ is modeled by considering three strategical
types of players: ({\it i\/}) cooperators \C\ and ({\it ii\/}) defectors \D\ both
willing to join the \PGG, with different intentions though. While the former are
ready to contribute a fixed share to the common pool, the latter attempt to exploit
the resource. Finally there are the so-called ({\it iii\/}) loners \Lo\ which refuse
to participate and rather rely on some small but fixed income.
The loner strategy is thus risk averse. These strategies
lead to a rock-scissors-paper dynamics with cyclic dominance: if cooperators abound,
they can be exploited by defectors, but if defectors prevail it is best to abstain
and if no one participates in the \PGG, small groups can form and it pays to return
to cooperation. Therefore, voluntary participation provides a simple yet natural
way to avoid deadlocks in states of mutual defection. In well-mixed populations,
i.e. in mean-field type models with replicator dynamics \cite{hofbauer:98},
this system can be solved analytically \cite{hauert:jtb02}.

In this letter, we consider a spatially extended variant of the voluntary \PGG\
where players are arranged on a rigid regular lattice and interact with their local
neighborhood only. Each player is confined to a site \x\ on a square lattice.
The size of the neighborhood therefore determines the maximum number of
participants $N$ in the \PGG. We restrict our investigations to the
\emph{von Neumann} neighborhood, i.e. to $N=5$. But note that the qualitative
results remain unaffected by the underlying geometry of the regular lattice.
The state variable \sx$\in\{$\C,\D,\Lo$\}$ determines the player's strategy at
any given time. The score achieved in \PGG\ interactions denotes the reproductive
success, i.e. the probability that one of the neighbors will adopt the player's
strategy. In the rigorous sense of the spatial \PGG, this score is accumulated
over $N=5$ games, i.e. by summing up the player's performance in \PGG s taking
place on the player's site as well as on the neighboring sites. For the sake of
simplicity, we assume that the score \Px\ is determined by a single, typical
\PGG\ involving the player and its four nearest neighbors. This simplification
accelerates the simulations and makes the pair approximation more convenient
while causing minor modifications in the system's dynamics.

The score \Px\ depends on the five strategies. Namely, if $n_c$, $n_d$, and
$n_l$ (with $n_c+n_d+n_l=N=5$) denote the number of participants choosing \C, 
\D\ and \Lo, then
\begin{equation}
P({\bf x})= 
\left\{		% make bracket-matching editors happy... }
\begin{array}{ll}
   \lfrac{r n_c}{n_c+n_d}-1 & \text{if \sx\ $=$ \C ,} \vspace{2mm}\\
   \lfrac{r n_c}{n_c+n_d}   & \text{if \sx\ $=$ \D ,} \vspace{2mm}\\ 
   \sigma		    & \text{if \sx\ $=$ \Lo ,}
\end{array}
\right.
\end{equation}
where the cooperative investments are normalized to unity and $r$ specifies the
multiplication factor on the public good. Note that $r>1$ must hold such that
groups of cooperators are better off than groups of defectors - hence to establish
a social dilemma. The loner payoff $\sigma$ with $0<\sigma<r-1$ denotes a small
but reliable source of income with a lower performance than mutual cooperation
but better than mutual defection. Solitary \C\ or \D\ players ($n_c+n_d=1$)
are assumed to act as loners.

Players reassessing and updating their strategies are randomly chosen (e.g. at
site \x) and compare their score to a randomly chosen neighbor \y. \x\ adopts the
strategy of \y\ with a probability \cite{szabo:pre98}:
\begin{equation}
\label{eq:W}
W[s({\bf y})\to s({\bf x})] = 
\frac{1}{1+\exp[(P({\bf x})-P({\bf y})+\tau)/K]}
\end{equation}
where $\tau>0$ denotes the cost of strategy change and $K$ introduces some noise
to allow for irrational i.e. non-payoff-maximizing choices. For $K=0$ the
neighboring strategy $s({\bf y})$ is always adopted provided the payoff difference
exceeds the cost of strategy change, i.e. $P({\bf y})>P({\bf x})+\tau$. For $K>0$,
strategies performing worse are also adopted with a certain probability e.g. due
to imperfect information. $K$ determines the half-width of this probability
distribution.

By means of Monte Carlo (\MC) simulations complemented by pair approximation, we
determine the equilibrium frequencies of the three strategies when varying $r$
while keeping $\sigma, K$ and $\tau$ fixed. For the pair approximation we determine
analytically the doublet density i.e. the probability of all configurations of two 
neighboring sites \cite{szabo:pre98}. Through moment closure i.e. by approximating 
higher order densities (e.g. triplets) with doublet densities, a set of equations of motion 
is obtained which is solved numerically.

Qualitatively the dynamics remains unaffected when changing $\sigma, K$ and 
$\tau$ within realistic limits. Henceforth we thus concentrate on the general features 
of spatio-temporal patterns and transitions. As we shall see, the cyclic dominance of 
the strategies acts as a driving force for traveling waves and leads to persistent and 
robust co-existence of all three strategies over a wide parameter range. Similar 
results have been found for an externally driven variant of the spatially extended
\PD\ with three strategies \cite{szabo:pre00} or if sites are allowed to go
empty \cite{nowak:jbifchaos94}. The simulations are performed under periodic
boundary conditions on an $M \times M$ lattice with $400\leq M\leq 2000$.
In general, we choose a random initial state and after suitable thermalization
times we determine the average frequency and fluctuation of the three strategies.

Let us first briefly consider the compulsory \PGG, i.e. with \C\ and \D\ only.
The spatial extension may enable cooperators to persist by forming clusters and
thereby minimizing exploitation by defectors. This is a well-known result from
other cooperation games \cite{nowak:nature92:space,szabo:pre98,hauert:jbifchaos02}.
For sufficiently high $r>r_{\text{C}}$ cooperators survive with frequencies
quickly increasing with $r$ because \C\ is favored for an increasing number of
local configurations. In contrast, below the threshold $r_{\text{C}}$ the
system eventually reaches the homogeneous \D\ state (see Fig.~1). Henceforth,
the subscript $\alpha$ of $r_{\alpha}$ refers to the vanishing strategy.
\begin{figure}
%\centerline{\epsfig{file=dc_r.pdf,width=8cm}}
\centerline{\epsfig{file=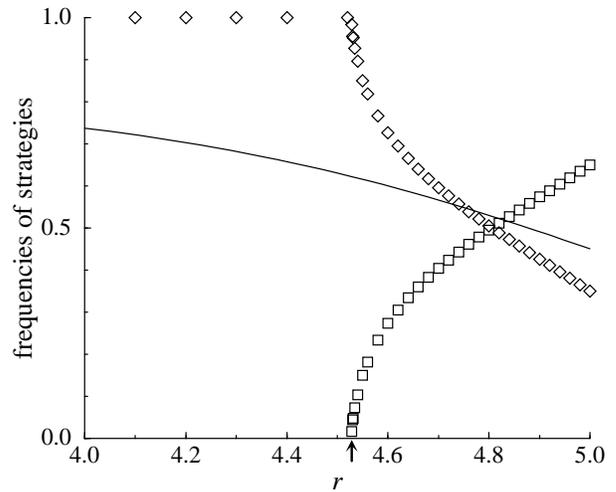,width=8cm}}
\caption{\label{fig:dc_r}Frequencies of cooperators \C\ (open squares) and
defectors \D\ (open diamonds) as a function of the multiplication rate $r$ for 
$\sigma=1$ and $\tau=K=0.1$. The solid line shows the frequency of defectors
in pair approximation. The arrow indicates $r_{\text{C}}$ where cooperators 
vanish.}
\end{figure}

In the close vicinity of $r_{\text C}$, the visualization of strategy distribution 
shows isolated colonies of $C$. These colonies move randomly and can coalesce
or divide. Consequently, this system becomes equivalent to a branching and
annihilating random walk \cite{cardy:jsp98} which exhibits a transition belonging 
to the directed percolation (DP) universality class \cite{DPconj,DPexp,szabo:pre98,chiappin:pre99}.
According to \MC\ simulations for $r\to r_{\text{C}}$ from above, the frequency
of \C\ is proportional to $(r-r_{\text{C}})^\beta$ with $r_{\text{C}}=4.526(1)$
and $\beta=0.55(3)$ for $\sigma=1$ and $K=\tau=0.1$. The pair approximation 
predicts a significantly lower critical value $r_{\text{C}}^{(p)}=2.694$. This 
difference refers to the enhanced role of $n$-point ($n>2$) correlations. The 
four-point approximation is expected to yield more accurate results \cite{szabo:pre00}.

In the case of voluntary participation, the loners induce significant changes
most pronounced at low $r$. The resulting dynamics can be divided into three
regimes (see Fig.~2):
\begin{figure}
%\centerline{\epsfig{file=c_r.pdf,width=8cm}}
\centerline{\epsfig{file=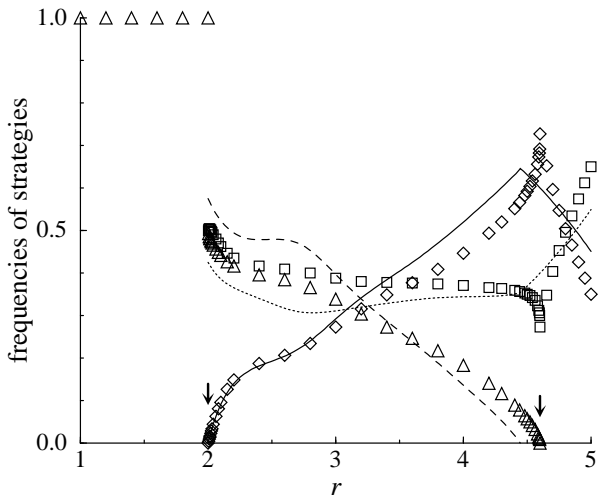,width=8cm}}
\caption{\label{fig:c_r}Frequencies of cooperators \C\ (open squares), defectors
\D\ (open diamonds) and loners \Lo\ (open triangles) as a function of $r$
for $\sigma=1$ and $\tau=K=0.1$. The results of pair approximation are shown
as dotted (\C), solid (\D) and dashed (\Lo) lines. The values of $r_{\text{D}}$
and $r_{\text{L}}$ are indicated by arrows.}
\end{figure}
(a) For $r<r_{\text{D}}=1+\sigma$ it is trivial that only loners survive
since they perform better than groups of cooperators. Note that solitary
\C\ and \D\ are eliminated by noise. (b) For $r_{\text{D}}<r<r_{\text{L}}$
the three strategies co-exist and produce fascinating spatio-temporal
patterns including traveling waves \cite{hauert:virtualpgg02}.
Such values of $r$ almost invariably result in homogeneous \D\ states in
the compulsory \PGG. Thus, the loners provide vital
protection to cooperators against exploitation. (c) For $r>r_{\text{L}}$
cooperators again thrive on their own as in the compulsory \PGG. Loners go
extinct because they no longer provide a valuable alternative.

In the remaining text we discuss the co-existence regime in greater detail.
According to our numerical analysis, the extinction of loners for
$r\to r_{\text{L}}$ also exhibits a \DP\ transition
\cite{DPconj,marro:99,hinrichsen:advphys00}. The frequency of \Lo\ is
proportional to $(r_{\text{L}}-r)^\beta$ in the vicinity of $r_{L}=4.6005(5)$
with $\beta=0.58(3)$ in agreement with previous data \cite{DPexp}. The increase
of fluctuations in the frequency of loners is consistent with a power law
divergence predicted by scaling hypothesis \cite{marro:99,hinrichsen:advphys00}.

The robustness of \DP\ transitions is well demonstrated by noting that the two
critical transitions belong to the same universality class despite remarkable
differences. The extinction of \C, leaving a homogeneous \D\ state behind,
contrasts with the extinction of \Lo\ on a time-dependent, inhomogeneous
\C$+$\D\ background. Field-theoretic arguments indicate that the main features
of \DP\ remain unchanged if the spatio-temporal fluctuations of the random
environment are uncorrelated \cite{hinrichsen:advphys00}. Our numerical results
support this expectation.

In the region of co-existence, the frequency of \C\ remains within narrow
limits compared to the trends observed for \D\ and \Lo. Figure~2 indicates
that the pair approximation yields a suitable quantitative description.
In particular, \D\ vanishes linearly with $r\to r_{\text{D}}$. This
behavior is strongly related to pattern evolution observed for low
\D\ frequency (see Fig.~3).
\begin{figure}
%\centerline{\epsfig{file=snapshot.pdf,width=8cm}}
\centerline{\epsfig{file=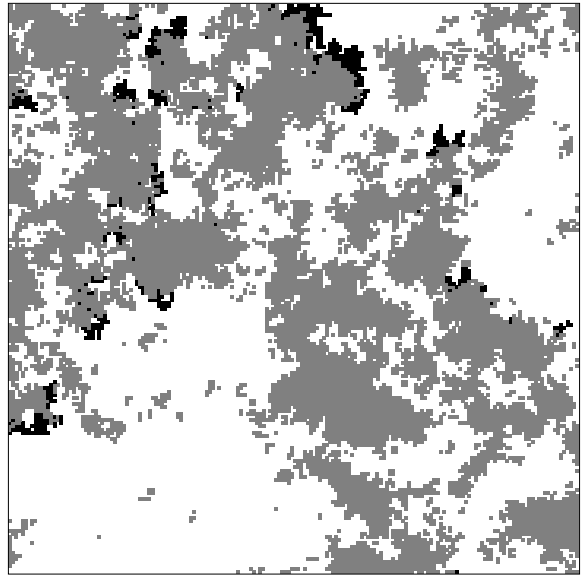,width=8cm}}
\caption{\label{fig:snapshot}Distribution of cooperators (white), defectors
(black) and loners (gray) on a $200\times 200$ portion of a larger lattice.
The parameters are $\sigma=1, \tau=K=0.1$ and $r=2.035$ i.e. slightly above
the transition to homogeneous states of loners.}
\end{figure}
The \D\ strategy forms small black islands invading the territory of \C. At the
same time, defectors are in turn invaded by loners paving the way for the return
of cooperators. The cyclic dominance maintains this self-organizing pattern.
But defectors can easily die out if the system size is not large enough.
The occasional extinction of \D\ results in a homogeneous \C\ state. Therefore,
this requires extremely large system sizes and a careful preparation of the
initial state. For example, $r$ may be gradually decreased until the desired
value is reached, or an artificial initial state may be prepared with neighboring
strips of \D\ and \Lo\ in a world of \C. Interestingly, the fluctuations of
the \D\ frequency remains constant while the frequency itself vanishes linearly
(see Fig.~4).
\begin{figure}
%\centerline{\epsfig{file=chi_r.pdf,width=8cm}}
\centerline{\epsfig{file=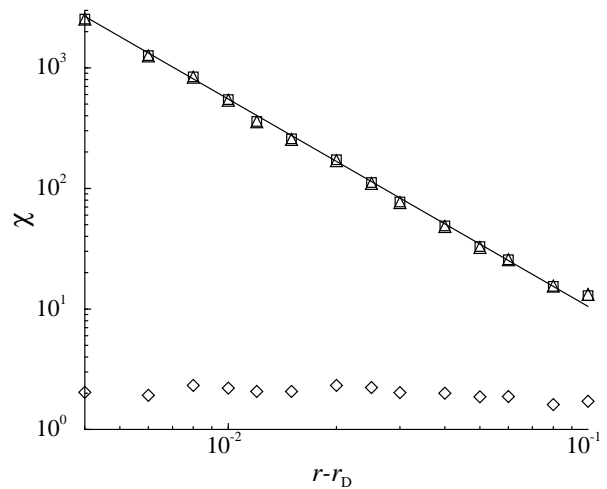,width=8cm}}
\caption{\label{fig:chi_r}Log-log plot of the frequency fluctuations $\chi$ vs.\
$r-r_{\text{D}}$ (symbols and parameters as in Fig.~2). $\chi$ denotes the square of
average fluctuation amplitudes produced by the system of size $M^2$. The solid line
shows the fitted power law with $\gamma=1.72$.}
\end{figure}
Consequently, to avoid accidental extinctions of \D, sufficiently large system
sizes $M$ are required such that the average number of \D\ is much larger than
the root-mean-square of their fluctuations. 

The typical domain size increases when $r$ goes to $r_{\text{D}}$. Figure 4
illustrates that this is accompanied by a power law divergence of the frequency 
fluctuations for \C\ and \Lo\ strategies 
$\chi_c\simeq\chi_l\propto(r-r_{\text{D}})^{-\gamma}$. The numerical fit
gives an exponent close to $\gamma=7/4$ which is characteristic to the
order parameter fluctuations in the Ising model when approaching the critical
point from above \cite{stanley:71}. One might argue that the multiplication factor 
$r$ is related to an external field stimulating cooperation and the noise term $K$ 
to temperature, however, a direct mapping seems impossible due to the additional
dependence on the loner's payoff $\sigma$.

Another curiosity of this model refers to the equal frequencies of \C\ and \Lo\
in this limit. Moreover, the correlation length $\xi$ (derived from the
density-density correlation function, see Fig.~5) appears to be
proportional to $1/(r-r_{\text{D}})$. The formation of larger and larger
domains in the two-dimensional, zero-field Ising model exhibits similar
behavior when decreasing the temperature to the critical point
\cite{stanley:71}. This suggests that the universality of this Ising type
transition determines constraints on the size distributions of \C\ and
\Lo\ domains. In this case the defectors with vanishing frequency contribute
to maintain suitable domain dynamics.
\begin{figure}
%\centerline{\epsfig{file=cl_r.pdf,width=8cm}}
\centerline{\epsfig{file=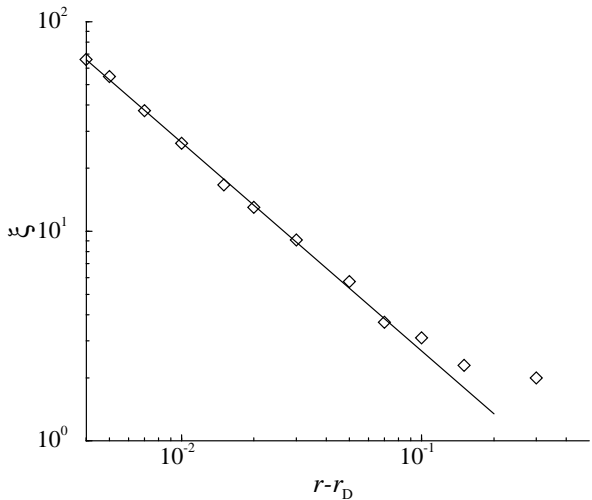,width=8cm}}
\caption{\label{fig:cl_r}Log-log plot of the correlation length $\xi$ vs.\
$r-r_{\text{D}}$ for the three strategies (parameters as in Fig.~2).
The solid line denotes the fitted power law with an exponent $-0.99$.}
\end{figure}

To conclude, we introduced a spatial evolutionary \PGG\ model demonstrating
that the successful spreading of selfish behavior is efficiently prevented
by allowing for voluntary participation. In the compulsory \PGG, i.e. in
absence of loners, cooperators thrive only if clustering advantages are
strong enough which requires sufficiently high multiplication factors $r$.
The introduction of loners leads to a cyclic dominance of the strategies
and promotes substantial levels of cooperation where otherwise defectors
dominate.

\begin{acknowledgments}
This work was supported by the Hungarian National Research Fund under
Grant No. T-33098. Ch.~H. acknowledges support of the Swiss National Science
Foundation.
\end{acknowledgments}

%\nocite{janssen:zpb81,grassberger:zpb82,brower:pl78,jensen:pra90}
%\bibliography{game,physics}

% the following mess is created by bibtex - 
% few things changed to have multiple references under one number
% see DPconj and DPexp
%

\end{document}